\documentstyle[12pt,rotate]{article}

\input epsf.sty

\newcommand{\ft}[2]{{\textstyle\frac{#1}{#2}}}

\newcommand{\g}{{\sl g}}

\newcommand{\cD}{{\cal D}}

\newcommand{\cN}{{\cal N}}

\newcommand{\cL}{{\cal L}}

\newcommand{\mG}{{\mit\Gamma}}
\newcommand{\mP}{{\mit\Psi}}
\newcommand{\mS}{{\mit\Sigma}}
\newcommand{\phan}[1]{{\textstyle\phantom{#1}}}
\font\cmss=cmss12 
\def\1{\hbox{{1}\kern-.25em\hbox{l}}}
\def\bfZ{\relax{\hbox{\cmss Z\kern-.4em Z}}}

\begin{document}
\begin{titlepage}

\centerline{\large \bf Instantons, Euclidean supersymmetry
                       and Wick rotations. }

\vspace{15mm}

\centerline{\bf A.V. Belitsky, S. Vandoren and P. van Nieuwenhuizen}

\vspace{15mm}

\centerline{\it C.N.\ Yang Institute for Theoretical Physics}
\centerline{\it State University of New York at Stony Brook}
\centerline{\it NY 11794-3800, Stony Brook, USA}

\vspace{30mm}

\centerline{\bf Abstract}

\hspace{0.5cm}

We discuss the reality properties of the fermionic collective coordinates
in Euclidean space in an instanton background and construct hermitean
actions for $\cN = 4$ and $\cN = 2$ supersymmetric Euclidean Yang-Mills
theories.

\vspace{7cm}


\vspace{0.5cm}


\end{titlepage}

\section{Introduction.}

In Euclidean supersymmetric Yang-Mills field theories (SYM)
\cite{Zum77,Nic78,Nie96}, the choice between an instanton or
anti-instanton background determines whether the chiral or anti-chiral
fermions acquire zero mode solutions parametrized by Grassmann collective
coordinates. In order that the action written in two-component formalism
with Weyl spinors $\lambda^\alpha$ and $\bar\lambda_{\dot\alpha}$ be
hermitean, we need a reality condition on the fermions in Euclidean space;
however, the latter is usually not specified. If we were to continue using
the same action with spinors $\lambda^\alpha$ and $\bar\lambda_{\dot\alpha}$
as in Minkowski space we run immediately into a problem: if $\lambda_\alpha$
contains fermionic zero modes in the presence of an anti-instanton,
$\bar\lambda_{\dot\alpha}$ does not have any zero modes, and this cannot be
reconciled with the Majorana condition $\bar\lambda_{\dot\alpha} = \left(
\lambda_\alpha \right)^\dagger$. If one just would forget about this
condition on the fermions, one would find a similar problem for the bosons
at the level of field equations of motion. For example, in the $\cN = 4$
supersymmetric model $\cite{Sch77}$, the six scalars $\phi^{AB}$ in the
${\bf 6}$ of $SU(4)$ are usually taken to satisfy the condition
\begin{equation}
\label{MinkRealScal}
\left( \phi^{AB} \right)^* \equiv \bar\phi_{AB} = \ft12
\epsilon_{ABCD} \phi^{CD} ,
\end{equation}
and their field equation in Minkowski space-time reads
\begin{equation}
\cD^2 \phi^{AB}
+ \sqrt{2} \left\{ \lambda^{\alpha, A}, \lambda_{\alpha}^B \right\}
+ \ft1{\sqrt{2}} \epsilon^{ABCD}
\left\{ \bar\lambda^{\dot\alpha}_C, \bar\lambda_{\dot\alpha, D} \right\}
- \ft12 \left[ \bar\phi_{CD}, \left[ \phi^{AB},
\phi^{CD} \right] \right] = 0 .
\end{equation}
But if one were to use the same field equation in Euclidean space
with an instanton solution (only changing $\ft{d}{dt}$ and $A_0$ into
$i \ft{d}{d\tau}$ and $i A_4$, respectively) one would find that since
only $\lambda^{\alpha,A}$ carries fermionic zero modes, the reality
condition in (\ref{MinkRealScal}) is violated. Some physicists are
confident that, as far as explicit calculations are concerned, no problem
arises if one views all fields as complex without reality conditions. (The
action is holomorphic: it depends on fields, not their complex conjugates.
For fermions an independent $\bar\psi$ or a dependent $\bar\psi =
(\psi)^*$ gives the same Grassmann integration \cite{Nie96}.)

Our resolution of this problem is consistent with both points of view:
the action contains complex fields and is the same in Euclidean and
Minkowski spaces, but consistent reality conditions for the fields
exist in Euclidean space and are different from those in Minkowski
space. These reality conditions involve both the space-time and internal
$R$-symmetry groups. To derive the correct reality conditions on
spinors and scalars such that the resulting action is hermitean, we
use dimensional reduction on a torus with one time coordinate from
the $\cN = 1$ SYM in $(9,1)$ or $(5,1)$ Minkowski space to $(4,0)$
Euclidean space with 16 and 8 supercharges, respectively \cite{Bla97}.
We shall see that in the reality condition of fermions the ``space-time
metric'' $\gamma^4$ is replaced by a metric $\eta_{AB}$ of the $R$-symmetry
group. This metric appears since the compact $R$ symmetry group of
the $\cN = 4$ Minkowski model becomes non-compact when the non-compact
space-time group $SO(3,1)$ is converted to the compact Euclidean group
$SO(4)$. We discuss in the following two sections both the $\cN = 4$
and the $\cN = 2$ model and will comment on the $\cN = 1$ model in the
conclusion.

\section{$\cN = 4$ Euclidean model.}

To construct the $\cN = 4$ supersymmetric YM model in Euclidean $d =
(4, 0)$ space, we start with the $\cN = 1$ SYM model in $d = (9, 1)$
Minkowski space-time, and reduce it on a six-torus with one time and
five space coordinates. In this way the Euclidean action and reality
conditions on Euclidean fields are automatically produced and can be
compared with their Minkowski counterparts \cite{Sch77} by reducing the
same $\cN = 1$ theory on a torus with six space coordinates. This
reduction is expected to lead to an internal non-compact $SO(5, 1)$
symmetry group in Euclidean space which is the Wick rotation of the
$SU(4) = SO (6)$ $R$-symmetry group in $d = (3, 1)$. The reality conditions
on bosons and fermions will both use an internal metric for this non-compact
internal symmetry group, and since for the instanton solution one uses
the 't Hooft symbols $\eta^a_{\mu\nu}$ (self-dual) and
$\bar\eta^a_{\mu\nu}$ (anti-self-dual) it is natural to use them also
for the internal metric. For spinors the internal metric follows by
dimensional reduction from the properties of the Dirac matrices in $10$,
$6$ and $4$ dimensions, hence we use $\eta$ and $\bar\eta$ symbols in
the construction of Dirac matrices in $6$ dimensions. In $4$ dimensions
we take the usual off-diagonal representation in terms of $\sigma_\mu$
and $\bar\sigma_\mu$, but note that all four $\gamma_\mu$ are hermitean
and square to $+1$. One of the Dirac matrices in $6$ dimensions will
be associated to the time coordinate and has square $- 1$; all other
are again hermitean with square $+ 1$. We shall need properties of the
charge conjugation matrices in $d = (9, 1)$, $d = (5, 1)$ and $d =
(4, 0)$. It has been shown by means of finite group theory \cite{Nie81}
that all these properties are representation independent. In particular,
there are two charge conjugation matrices $C^+$ and $C^-$ in even
dimensions, satisfying $C^\pm \mG^\mu = \pm \left( \mG^\mu \right)^T
C^\pm$, and $C^+ = C^- \ast\!\mG$, where $\ast\!\mG$ is the product of
all Dirac matrices normalized to $\left( \ast\!\mG \right)^2 = + 1$.
Then $\ast\!\mG$ is a hermitean matrix. These charge conjugation
matrices do not depend on the signature of space-time and
$C^- \ast\!\mG = \pm \left(\ast\!\mG\right)^T C^-$ with $-$ sign in $d
= 10, 6$ but with $+$ sign in $d = 4$. In $d = 10$ $\left( C^\pm
\right)^T = \pm C^\pm$ while for $d = 6$ $\left( C^\pm \right)^T =
\mp C^\pm$, finally for $d = 4$ $\left( C^\pm \right)^T = - C^\pm$.

We start with the $d = (9,1)$ $\cN = 1$ Lagrangian
\begin{equation}
\cL_{10} = \frac{1}{\g^2_{10}} {\rm tr}
\left\{
\ft12 F_{MN} F^{MN} + \bar \mP \mG^M \cD_M \mP
\right\} ,
\end{equation}
where $F_{MN} = \partial_M A_N - \partial_N A_M + \left[ A_M, A_N \right]$
and $\mP$ is a Majorana-Weyl spinor
\begin{equation}
\label{MajoranaWeyl}
\mG^{11} \mP = \mP,
\qquad
\mP^T C^-_{10} = \mP^\dagger i \mG^0 \equiv \bar \mP ,
\end{equation}
where $\mG^{11} \equiv \ast\!\mG$. Both $A$ and $\mP$ are Lie algebra
valued, $A = A^a T_a$ and $\mP = \mP^a T_a$ with anti-hermitean $SU(N)$
generators $T^a$ normalized according to ${\rm tr} T^a T^b = - \ft12
\delta^{ab}$. The $\mG$-matrices obey the Clifford algebra $\left\{
\mG^M, \mG^N \right\} = 2 \eta^{MN}$ with metric $\eta^{MN} = {\rm diag}
(-, +, \dots, +)$. The Lagrangian is a density under the following
transformation rules
\begin{equation}
\delta A_M = \bar \zeta \mG_M \mP ,
\qquad
\delta {\mit\Psi} = - \ft12 F_{MN} \mG^{MN} \zeta,
\end{equation}
with $\mG^{MN} = \ft12 [\mG^M \mG^N - \mG^M \mG^N]$ and $\bar\zeta =
\zeta^T C^-_{10}$. (Of course, $\zeta^T C^-_{10} = \zeta^\dagger i \mG^0$.)

\begin{table}
\begin{center}
\begin{tabular}{lll}
\hline
 & & \\
$\gamma^\mu
= \left(
\begin{array}{cc}
0                              & - i \sigma^{\mu,\alpha\beta'} \\
i \bar\sigma^\mu_{\alpha'\beta} & 0
\end{array}
\right)$,
&
$\gamma^5 = \gamma^1 \gamma^2 \gamma^3 \gamma^4
= \left(
\begin{array}{cc}
1 & 0 \\
0 & - 1
\end{array}
\right)$,
& $C_4^- = \gamma^4 \gamma^2
= \left(
\begin{array}{cc}
\epsilon_{\alpha\beta} & 0                      \\
0                      & \epsilon^{\alpha'\beta'}
\end{array}
\right)$ \\
& & \\
$\hat\gamma^a
= \left(
\begin{array}{cc}
0               & \mS^{a, AB} \\
\bar\mS^a_{AB} & 0
\end{array}
\right)$,
& $\hat\gamma^7 = \hat\gamma^1 \dots \hat\gamma^6
= \left(
\begin{array}{cc}
1 & 0 \\
0 & - 1
\end{array}
\right)$,
& $C_6^- = i \hat\gamma^4 \hat\gamma^5 \hat\gamma^6
= \left(
\begin{array}{cc}
0                  & \delta_{A}^{\phan{A} B}  \\
\delta^{A}_{\phan{A} B} & 0
\end{array}
\right)$ \\
 & & \\
\hline
\end{tabular}
\end{center}
\caption{\label{ten} \small The Dirac matrices in four and six dimensions
and corresponding charge conjugation matrices \protect\cite{DorHolKhoMatVan99}.
Here $\sigma^\mu = ( \vec\sigma, i)$ and $\bar\sigma^\mu = ( \vec\sigma,
- i)$ for $\mu = 1, \dots, 4$, and $\mS^{a, AB} = \left\{ - i \eta^{1,AB},
\eta^{2,AB}, \eta^{3,AB}, i \bar\eta^{k,AB} \right\}$, $\bar\mS^a_{AB}
= \left\{ i \eta^1_{AB}, - \eta^2_{AB}, - \eta^3_{AB}, i \bar\eta_{AB}^k
\right\}$ are expressed in terms of 't Hooft symbols \protect\cite{tHo76}.
Numerically, $\eta^{a,AB} = \eta^a_{AB}$ and the same for $\bar\eta$.
Furthermore, $\epsilon_{\alpha\beta} = - \epsilon^{\alpha'\beta'}$,
$\epsilon_{\alpha\beta} = \epsilon^{\alpha\beta}$.} 
\end{table}

To proceed with the dimensional reduction we choose a particular
representation of the gamma matrices in $d = (9, 1)$ , namely\footnote{Use
$\eta^a_{AB} \eta^b_{BC} = - \delta^{ab} \delta_{AC} - \epsilon^{abc}
\eta^c_{AC}$. The same relation holds for $\bar\eta^a_{AB}$. Further,
$[\eta^a, \bar\eta^b] = 0$.},
\begin{equation}
\label{10Gammas}
\mG^M
= \left\{ \hat\gamma^a \otimes \gamma^5, \1_{[8 \times 8]}
\otimes \gamma^\mu \right\},
\qquad
\mG^0
= \left(
\begin{array}{cc}
0                  & - i \eta^1  \\
i \eta^1           & 0
\end{array}
\right) \otimes \gamma^5 ,
\qquad
\mG^{11} = \mG^0 \dots \mG^9 = \hat\gamma^7 \otimes \gamma^5 ,
\end{equation}
where the $8 \times 8$ Dirac matrices $\hat\gamma^a$ and $\hat\gamma^7$
of $d = (5,1)$ with $a = 1, \dots, 6$ and the usual $\gamma^\mu$ and
$\gamma^5$ of $d = (4, 0)$ are given in the Table \ref{ten}. The charge
conjugation matrix $C^-_{10}$ is given by $C_{6}^- \otimes C_{4}^-$. Upon
compactification to Euclidean $d = (4, 0)$ space the 10-dimensional
Lorentz group $SO(9,1)$ reduces to $SO(4) \times SO(5,1)$ with compact
space-time group $SO(4)$ and $R$-symmetry group $SO(5,1)$. In these
conventions a $32$-component chiral Weyl spinor $\mP$ decomposes as follows
into $8$ and $4$ component chiral-chiral and antichiral-antichiral
spinors
\begin{equation}
\mP
= \left(
\begin{array}{c}
1 \\
0
\end{array}
\right)
\otimes
\left(
\begin{array}{c}
\lambda^{\alpha, A} \\
0
\end{array}
\right)
+
\left(
\begin{array}{c}
0 \\
1
\end{array}
\right)
\otimes
\left(
\begin{array}{c}
0 \\
\bar\lambda_{\alpha', A}
\end{array}
\right) ,
\end{equation}
where $\lambda^{\alpha, A}$ ($\alpha = 1,2$) transforms only under
the first $SU(2)$ in $SO(4) = SU(2) \times SU(2)$, while
$\bar\lambda_{\alpha', A}$ changes only under the second $SU (2)$.
Furthermore, $\bar\lambda_{\alpha',A}$ transforms in the complex
conjugate of the $SO(5,1)$ representation of $\lambda^{\alpha,A}$,
namely, $\left( \lambda^\ast \right)^{\alpha,B} \eta^1_{BA}$
transforms like $\bar\lambda_{\alpha,A}$, and the two spinor
representation of $SO(5,1)$ are pseudoreal, i.e.\ $[\hat\gamma^a ,
\hat\gamma^b]^\ast_L \eta^1 = \eta^1 [\hat\gamma^a , \hat\gamma^b]_R$
where $L$ ($R$) denotes the upper (lower) $4$-component spinor. The
Majorana condition (\ref{MajoranaWeyl}) on $\mP$ leads in Euclidean
space to reality conditions\footnote{Unless specified otherwise,
equations which involve hermitean or complex conjugation of fields
will be understood as not Lie algebra valued, i.e.\ they hold for
the components $\lambda^{a, \alpha, A}$, etc.} on $\lambda^\alpha$
which are independent of those on $\bar\lambda_{\alpha'}$, namely,
\begin{equation}
\label{RealityConds1}
\left( \lambda^{\alpha, A} \right)^\dagger
= - \lambda^{\beta, B} \epsilon_{\beta\alpha} \eta^1_{BA} ,
\qquad
\left( \bar\lambda_{\alpha', A} \right)^\dagger
= - \bar\lambda_{\beta', B} \epsilon^{\beta'\alpha'} \eta^{1, BA} .
\end{equation}

These reality conditions are consistent \cite{Nie81} and define a
symplectic Majorana spinor in Euclidean space. The $SU (2) \times SU (2)$
covariance of (\ref{RealityConds1}) is obvious from the pseudoreality of
the ${\bf 2}$ of $SU (2)$, but covariance under $SO(5,1)$ can also be
checked (use $[\eta^a, \bar\eta^b] = 0$).

Substituting these results, the action reduces to
\begin{eqnarray}
\label{N4susy}
\cL_E^{\cN = 4}\!\!\! &=&\!\!\! \frac{1}{\g^2} {\rm tr}\
\Bigg\{
\ft12 F_{\mu\nu} F_{\mu\nu}
- 2 i \bar\lambda^{\alpha'}_{A} \bar\sigma_{\mu, \alpha'\beta}
\cD_\mu \lambda^{\beta, A}
+ \ft12 \left( \cD_\mu \bar\phi_{AB} \right)
\left( \cD_\mu \phi^{AB} \right)
\nonumber\\
&-&\!\!\! \sqrt{2}
\bar\phi_{AB} \left\{ \lambda^{\alpha, A}, \lambda_{\alpha}^{B} \right\}
- \sqrt{2}
\phi^{AB}
\left\{ \bar\lambda^{\alpha'}_{A}, \bar\lambda_{\alpha', B} \right\}
+ \ft18 \left[ \phi^{AB}, \phi^{CD} \right]
\left[ \bar\phi_{AB}, \bar\phi_{CD} \right]
\Bigg\} ,
\end{eqnarray}
with $\bar\phi_{AB} \equiv \ft12 \epsilon_{ABCD} \phi^{CD}$. The
anti-symmetric scalar $\phi^{AB}$ is defined by $\frac{1}{\sqrt{2}}
\mS^{a, AB} A_a $ where $A_a$ are the first six real components of the
ten dimensional gauge field potential $A_M$. Since the first $\mS$ matrix
has an extra factor $i$ in order that $\left( \mG^0 \right)^2 = - 1$,
see (\ref{10Gammas}), the reality condition on $\phi^{AB}$ involves
$\eta^1_{AB}$
\begin{equation}
\label{RealityConds2}
\left( \phi^{AB} \right)^\ast
= \eta^1_{AC} \phi^{CD} \eta^1_{DB} .
\end{equation}
The Euclidean action in (\ref{N4susy}) is hermitean under the reality
conditions in (\ref{RealityConds1}) and (\ref{RealityConds2}).
In fact, one obtains the same action for the Minkowski case by reducing
on a torus with $6$ space coordinates, but then we find the reality
conditions in (\ref{MinkRealScal}) and the usual Majorana condition:
$\bar\lambda_{\dot\alpha,A} = \left( \lambda^A_{\alpha} \right)^\dagger$.

The action is invariant under the dimensionally reduced supersymmetry
transformation rules\footnote{In Euclidean space we use the notation
$\alpha'$ instead of Minkowskian $\dot\alpha$. Then $\left( \delta
\lambda^{\alpha,A} \right)^\ast \epsilon_{\dot\alpha\dot\beta} = - 
\delta \bar\lambda_{\dot\beta, A}$, but there is no similar relation 
between $\left( \delta \lambda^{\alpha,A} \right)^\ast$ and $\delta 
\bar\lambda_{\alpha',A}$ in Euclidean space because $SO(4) = SU(2) 
\times SU(2)$ but $SO(6)$ is simple.}
\begin{eqnarray}
\delta A_\mu
\!\!\!&=&\!\!\! - i {\bar\zeta}^{\alpha'}_A
\bar\sigma_{\mu,\alpha'\beta} \lambda^{\beta, A}
+ i \bar\lambda_{\beta', A} \sigma_\mu^{\alpha\beta'} \zeta^A_\alpha ,
\nonumber\\
\delta \phi^{AB}
\!\!\!&=&\!\!\! \sqrt{2} \Big( \zeta^{\alpha,A} \lambda^B_\alpha
- \zeta^{\alpha,B} \lambda^A_\alpha
+ \epsilon^{ABCD} \bar\zeta^{\alpha'}_C \bar\lambda_{\alpha',D} \Big),
\nonumber\\
\delta \lambda^{\alpha,A}
\!\!\!&=&\!\!\! - \ft12 (\sigma^{\mu\nu})^\alpha_{\phan{\alpha}\beta}
F_{\mu\nu} \zeta^{\beta,A}
- i \sqrt{2} \bar\zeta_{\alpha',B} \sigma_{\mu}^{\alpha\alpha'}
\cD_\mu \phi^{AB}
+ \left[ \phi^{AB}, \bar\phi_{BC} \right] \zeta^{\alpha,C} ,
\nonumber\\
\delta \bar\lambda_{\alpha',A}
\!\!\!&=&\!\!\! - \ft12
(\bar\sigma^{\mu\nu})_{\alpha'}^{\phan{\alpha'}\beta'}
F_{\mu\nu} \bar\zeta_{\beta',A}
+ i \sqrt{2} \zeta^{\alpha,B} \bar\sigma_{\mu,\alpha'\alpha}
\cD_\mu \bar\phi_{AB}
+ \left[ \bar\phi_{AB}, \phi^{BC} \right] \bar\zeta_{\alpha',C} ,
\end{eqnarray}
where $\sigma_{\mu\nu} \equiv \ft12 [\sigma_\mu \bar\sigma_\nu -
\sigma_\nu \bar\sigma_\mu ]$ is anti-self-dual, $\ft12
\epsilon_{\mu\nu\rho\sigma} \sigma_{\rho\sigma} = - \sigma_{\mu\nu}$,
and $\bar\sigma_{\mu\nu} \equiv \ft12 [ \bar\sigma_\mu \sigma_\nu
- \bar\sigma_\nu \sigma_\mu ]$ is self-dual, $\ft12
\epsilon_{\mu\nu\rho\sigma} \bar\sigma_{\rho\sigma} = \bar\sigma_{\mu\nu}$,
and $\ft12 \epsilon_{ABCD} \mS^{a,CD} = - \bar\mS^a_{AB}$.
Also these rules are the same as in Minkowski case, but with modified
reality conditions.

One of the main motivations for studying Euclidean supersymmetric theories
is that they allow one to compute non-perturbative instantons effects in
correlation functions. In many correlators the one-loop perturbative
corrections occur only as quantum determinants which are the products of
the non-zero modes of the quantum fluctuations and cancel due to
supersymmetry. In these cases the analysis can be limited to the study of
zero modes. The Euclidean equations of motion
\begin{eqnarray}
&& \cD_\nu F_{\nu\mu} - i
\left\{ \bar\lambda^{\alpha'}_A \bar\sigma_{\mu,\alpha'\beta} ,
\lambda^{\beta, A} \right\}
- \ft12 \left[ \bar\phi_{AB}, \cD_\mu \phi^{AB} \right] = 0,
\nonumber\\
&& \cD^2 \phi^{AB}
+ \sqrt{2} \left\{ \lambda^{\alpha, A}, \lambda_{\alpha}^B \right\}
+ \ft1{\sqrt{2}} \epsilon^{ABCD}
\left\{ \bar\lambda^{\alpha'}_C, \bar\lambda_{\alpha', D} \right\}
- \ft12 \left[ \bar\phi_{CD}, \left[ \phi^{AB}, \phi^{CD} \right] \right]
= 0,
\nonumber\\
&& \bar\sigma_{\mu, \alpha'\beta} \cD_\mu \lambda^{\beta, A}
+ i \sqrt{2} \left[ \phi^{AB}, \bar\lambda_{\alpha', B} \right] = 0 ,
\quad
\sigma_{\mu, \alpha\beta'} \cD_\mu \bar\lambda^{\beta'}_{A}
+ i \sqrt{2} \left[ \bar\phi_{AB}, \lambda_{\alpha}^{B} \right] = 0 ,
\end{eqnarray}
are consistent with the reality conditions in Eqs.\ (\ref{RealityConds1},
\ref{RealityConds2}). We can obtain a complete solution to the above
equations by first solving exactly the simplified field equations
$\cD_\nu F_{\nu\mu} = 0$, $\bar\sigma_{\mu, \alpha'\beta} \cD_\mu
\lambda^{\beta, A} = 0$, $\sigma_{\mu, \alpha\beta'} \cD_\mu
\bar\lambda^{\beta'}_{A} = 0$. Choosing the anti-instanton configuration
\begin{equation}
A^{a, \bar I}_\mu (x; x_0, \rho)
= 2 \frac{\rho^2 \eta^a_{\mu\nu}
(x - x_0)_\nu}{ (x - x_0)^2 \left( (x - x_0)^2 + \rho^2 \right)},
\end{equation}
there are only solutions for $\lambda$. From index theory it follows that
there are $4 \times 2 \times N$ fermionic zero modes in the background of
a $k = 1$ anti-instanton. Namely, there are $4 \times 4$ supersymmetric
and superconformal zero modes \cite{Shi83}
\begin{equation}
\lambda^{\alpha,A} = \ft12
\left( \sigma^{\mu\nu} \right)^\alpha_{\phan{\alpha} \beta}
\left( \xi^{\beta,A} - \bar\eta^A_{\beta'}
\sigma_\rho^{\beta\beta'} x^\rho
\right) F^{\bar I}_{\mu\nu} ,
\end{equation}
with Grassmann collective coordinates (GCC) $\xi$ and $\bar\eta$ which to
some extent are superpartners of center-of-mass coordinate $x_0$ and scale
$\rho$ of the anti-instanton $A^{\bar I}_\mu (x; x_0, \rho)$. The remaining
$4 \times 2 \times (N - 2)$ zero modes are given by (setting $x_0 = 0$)
\cite{Cor79,DorKhoMat96}
\begin{equation}
\label{MUmodes}
\left( \lambda^{\alpha,A} \right)_u^{\phan{u} v}
= \left[ \frac{\rho^2}{x^2 (x^2 + \rho^2)^3} \right]^{1/2}
\left( \mu^A_u (x^\alpha)^v + (x^\alpha)_u \bar\mu^{A, v} \right) ,
\end{equation}
where for fixed $\alpha$ and $A$, the $N$-component vectors $\mu^A_u$
and $\left( x^\alpha \right)^v$ are given by
\begin{equation}
\mu^A_u
= \left( \mu^A_1 , \dots , \mu^A_{N - 2} , 0 , 0 \right) , \qquad
(x^\alpha)^v
= \left( 0 , \dots , 0 , x^\mu \sigma_\mu^{\alpha\beta'} \right)
\quad\mbox{with}\quad N - 2 + \beta' = v .
\end{equation}
Further, $\left( x^\alpha \right)_u = \left( x^\alpha \right)^v
\epsilon_{vu}$ and $\bar\mu^{A, v}$ possesses also $N - 2$ nonvanishing
components.

Our reality conditions for spinors in (\ref{RealityConds1}) imply then
reality conditions for the Grassmann collective coordinates.
Straightforward substitution yields the following results
\begin{equation}
\label{GCCreality}
\left( \xi^{A, \alpha} \right)^\dagger
=
- \xi^{\beta, B} \epsilon_{\beta\alpha} \eta^1_{BA} , \quad
\left( \bar\eta^{A}_{\alpha'} \right) ^\dagger
= - \bar\eta^B_{\beta'} \epsilon^{\beta'\alpha'} \eta^1_{BA} , \quad
\left( \mu^A_u \right)^\dagger
= \bar\mu^{B, u} \eta^1_{BA} , \quad
\left( \bar\mu^{A, u} \right)^\dagger
= - \mu^{B}_u \eta^1_{BA} ,
\end{equation}
where we have used the involution of Pauli matrices in Euclidean
space $\left( \sigma_\mu^{\alpha\beta'} \right)^\ast = \sigma_{\alpha\beta'}$.
One can now make an expansion of the field equations in the number of
fermion fields and solve them order by order in the number of GCC. To
second order in GCC one must solve $\cD^2 \phi^{AB} + \sqrt{2} \left\{
\lambda^{\alpha,A}, \lambda^B_\alpha \right\} = 0$
\cite{DorKhoMat96,DorKhoMatVan98,DorHolKhoMatVan99}. Substitution of
these solutions into the $\cN = 4$ action $S = - \int d^4 x \cL$ 
(\ref{N4susy}) leads to an extra term, in addition to the standard 
one-anti-instanton action
$S^{\bar I} = \frac{8 \pi^2}{\g^2}$, which lifts all the fermionic
zero modes except the $\xi$ and $\bar\eta$
\cite{DorKhoMatVan98,DorHolKhoMatVan99}
\begin{equation}
\Delta S = \frac{\pi^2}{4 \g^2 \rho^2} \epsilon_{ABCD}
\left( \bar\mu^{u, A} \mu^B_u \right)
\left( \bar\mu^{v, C} \mu_v^D \right) .
\end{equation}
This term is hermitean w.r.t.\ the relations (\ref{GCCreality})
since $\epsilon_{ABCD} \eta^1_{AA'} \eta^1_{BB'} \eta^1_{CC'}
\eta^1_{DD'} = \left( {\rm det}\, \eta^1 \right) \epsilon_{A'B'C'D'}$
and ${\rm det}\, \eta^1_{AB} = 1$.

\section{$\cN = 2$ Euclidean model.}

Let us now address the $\cN = 2$ super-Yang-Mills Euclidean model
\cite{Sal74} deduced by dimensional reduction from the $d = (5, 1)$
$\cN = 1$ theory. The Lagrangian reads
\begin{equation}
\label{D6Lagr}
\cL_6 = \frac{1}{\g^2_6} {\rm tr}
\left\{
\ft12 F_{MN} F^{MN} + \bar \mP_i \mG^M \cD_M \mP^i
\right\} ,
\end{equation}
with the symplectic Majorana-Weyl condition
\begin{equation}
\label{SMW6}
\mG^7 \mP^i = \mP^i,
\qquad
\mP^{i, T} C^-_6 \epsilon_{ij} = {\mP^j}^\dagger i \mG^0 .
\end{equation}
The action is invariant under transformations
\begin{equation}
\delta A_M = \bar \zeta_i \mG_M \mP^i ,
\qquad
\delta {\mit\Psi}^i = - \ft12 F_{MN} \mG^{MN} \zeta^i ,
\qquad i = 1, 2 ,
\end{equation}
(use $\bar\mP_i \mG^M \zeta^i = - \bar\zeta_i \mG^M \mP^i$). We choose a
different representation of the Dirac matrices as compared to the previous
section, namely
\begin{equation}
\mG^M = \{ \tau^a \otimes \gamma^5,
\1_{[2 \times 2]} \otimes \gamma^\mu \},
\end{equation}
with $\tau^0 = - i \sigma^2$ and $\tau^1 = \sigma^1$. Then the charge
conjugation matrix is $C^-_6 = \mG^0 \mG^2 \mG^4 = i \sigma^2 \otimes
C^-_4$ and $*\!\mG = \mG^7 = \mG^5 \mG^4 \dots \mG^0 = \sigma^3 \otimes
\gamma^5$. From (\ref{D6Lagr}) the four-dimensional $R$-symmetry $U(2)$
is manifest: the fermion transforms as the $({\bf 4}, {\bf 2})$
of $SO(5,1) \times SU(2)$.

The procedure of the reduction via the time direction is completely
equivalent to the one discussed in the previous section with
the decomposition of the $d = (5,1)$ Lorentz group into $SO(5,1) \to SO(4)
\times SO(1,1)$. This gives
\begin{eqnarray}
\label{N2Eucl}
\cL_E^{\cN = 2} \!\!\!&=&\!\!\! \frac{1}{\g^2} {\rm tr} \
\Bigg\{
\ft12 F_{\mu\nu} F_{\mu\nu}
- 2 i \bar\lambda^{\alpha'}_i \bar\sigma_{\mu, \alpha'\beta}
\cD_\mu \lambda^{\beta, i}
+ \left( \cD_\mu S \right)^2
- \left( \cD_\mu P \right)^2
- \left[ P, S \right]^2
\nonumber\\
&-&\!\!\! \left( S - P \right)
\left\{ \lambda^{\alpha, i}, \lambda_{\alpha, i} \right\}
- \left( S + P \right)
\left\{ \bar\lambda^{\alpha', i}, \bar\lambda_{\alpha', i} \right\}
\Bigg\} ,
\end{eqnarray}
where $P \equiv A^0$ and $S \equiv A^1$. We use $\lambda^{\alpha,i}
\epsilon_{ij} \epsilon_{\alpha\beta}= \lambda_{\beta,j}$ and analogously
for $\bar\lambda^i_{\alpha'}$. As expected, the $U(1)$ of the $U(2)$ becomes
non-compact in complete agreement with the results of \cite{Zum77,Nie96}:
the automorphism of the supersymmetry algebra is $SO(1,1)$ so that Eq.\
(\ref{N2Eucl}) is invariant under scaling transformations $(S + P) \to
\varrho (S + P)$, $(S - P) \to \varrho^{-1} (S - P)$, $\lambda \to
\varrho^{1/2} \lambda$ and $\bar\lambda \to \varrho^{-1/2} \bar\lambda$.
The $SU(2)$ survives the reduction and remains compact. The $\cN = 2$
supersymmetry transformation is given by
\begin{eqnarray}
&&\delta A_\mu
= i \bar\zeta^{\alpha'}_i
\bar\sigma_{\mu,\alpha'\beta} \lambda^{\beta,i}
- i \zeta_{\alpha,i} \sigma_{\mu}^{\alpha\beta'} \bar\lambda^i_{\beta'},
\quad
\delta S = \zeta_{\alpha,i} \lambda^{\alpha,i}
- \bar\zeta_i^{\alpha'} \bar\lambda^i_{\alpha'},
\quad
\delta P = \zeta_{\alpha,i} \lambda^{\alpha,i}
+ \bar\zeta_i^{\alpha'} \bar\lambda^i_{\alpha'}, \nonumber\\
&& \delta \lambda^{\alpha,i}
= - \ft12 \left( \sigma_{\mu\nu} \right)^\alpha_{\phan{\alpha}\beta}
F_{\mu\nu} \zeta^{\beta,i}
- i \bar\zeta^i_{\alpha'} \sigma_\mu^{\alpha\alpha'}
\cD_\mu \left( S + P \right)
- \left[ P, S \right] \zeta^{\alpha,i}, \nonumber\\
&& \delta \bar\lambda^i_{\alpha'}
= - \ft12
\left( \bar\sigma_{\mu\nu} \right)_{\alpha'}^{\phan{\alpha}\beta'}
F_{\mu\nu} \bar\zeta^i_{\beta'}
- i \zeta^{\alpha,i} \bar\sigma_{\mu,\alpha'\alpha}
\cD_\mu \left( S - P \right)
+ \left[ P, S \right] \bar\zeta^i_{\alpha'} .
\end{eqnarray}
The Lagrangian (\ref{N2Eucl}) is hermitean w.r.t.\ the reality conditions
\begin{equation}
\left( \lambda^{\alpha,i} \right)^\dagger
= i \lambda^{\beta,j} \epsilon_{ji} \epsilon_{\beta\alpha}, \qquad
\left( \bar\lambda_{\alpha'}^i \right)^\dagger
= - i \bar\lambda_{\beta'}^j \epsilon_{ji} \epsilon^{\beta'\alpha'} ,
\end{equation}
stemming from the $d = 6$ constraints (\ref{SMW6}).

\section{Conclusions.}

The dimensional reduction of higher dimensional SYM theories via the
time direction naturally leads to Euclidean $\cN > 1$ supersymmetric
models with a hermitean action. The Lagrangians of the Euclidean models
and the SUSY transformation rules written in covariant form w.r.t.\ the
internal $R$ group do not change their form as compared to the Minkowskian
ones, however, the compact internal $R$-symmetry group becomes
non-compact in Euclidean space and the reality conditions on fields
involve different metrics. For fermions we end up with symplectic
Majorana conditions in $(4,0)$. When these results are translated to the
reality conditions on the Grassmann collective coordinates we obtain a
real effective action for the collective coordinates induced by instantons.

We could have used other representation for the $d = (9,1)$ Dirac
matrices, for example,
\begin{equation}
\mG^M = \{ \hat\gamma^a \otimes \1_{[4 \times 4]} ,
\hat\gamma^7 \otimes \gamma^\mu \}.
\end{equation}
In this representation $C^-_{10} = C^-_6 \otimes C^+_4$ but the reality
conditions in (\ref{RealityConds1}) are unchanged because $C^+_4 = C^-_4
\gamma^5$ while $\mG^0$ differs also by a factor $\gamma^5$.

Obviously, our modus operandi is not applicable to the $\cN = 1$ model.
In Euclidean space due to absence of real Dirac matrices the generators
of the supersymmetry are complex four-component spinors. Since the
internal $R$-symmetry group is Abelian for simple supersymmetry we
cannot impose the symplectic Majorana condition. In this situation one
accepts the idea of complexification of all fields \cite{Luk87} of the
theory without a reality condition (and loose hermiticity of the action).
If, however, one wants to preserve a real gauge field one is forced to
enhance the $\cN = 1$ supersymmetry to the $\cN = 2$ SUSY \cite{Zum77}.

\end{document}